# Soliton dynamics in a fractional complex Ginzburg-Landau model

Yunli Qiu[1], Boris A. Malomed[2], Dumitru Mihalache[3], Xing Zhu[1], Li Zhang[4], and Yingji He[1]*

[1] *School of Photoelectric Engineering, Guangdong Polytechnic Normal University, Guangzhou 510665, China*

[2] *Department of Physical Electronics, School of Electrical Engineering, Faculty of Engineering, and Center for Light-Matter Interaction, Tel Aviv University, Tel Aviv 69978, Israel*

[3] *Horia Hulubei National Institute for Physics and Nuclear Engineering, P.O. Box MG-6, RO-077125, Bucharest-Magurele, Romania*

[4] *School of Physics and Optoelectronic Engineering, Foshan University, Foshan 528000, China*

*Corresponding author: heyingji8@126.com

The general objective of the work is to study dynamics of dissipative solitons in the framework of a one-dimensional complex Ginzburg-Landau equation (CGLE) of a fractional order. To estimate the shape of solitons in fractional models, we first develop the variational approximation for solitons of the fractional nonlinear Schrödinger equation (NLSE), and an analytical approximation for exponentially decaying tails of the solitons. Proceeding to numerical consideration of solitons in fractional CGLE, we study, in necessary detail, effects of the respective Lévy index (LI) on the solitons' dynamics. In particular, dependence of stability domains in the model's parameter space on the LI is identified. Pairs of in-phase dissipative solitons merge into single pulses, with the respective merger distance also determined by LI.

# 1. INTRODUCTION

Based on the path-integral approach, a fractional generalization of the Schrödinger equation, developed in the framework of the fractional quantum and statistical mechanics, was proposed by Laskin [1]. Subsequently, fractional Schrödinger equations (FSEs) have drawn much interest in various areas of physics [2-10]. In particular, they apply to fields in fractional-dimension spaces and dynamics of particles with fractional spin. Due to issues with handling nonlocal operators that represent fractional derivatives and Laplacians, characterized by the respective Lévy index (LI) [11], and scarcity of relevant experimental results, the advancement in this area was slow. An essential step forward was made by Longhi [12], who has introduced FSE into optics and obtained solutions for dual Airy beams for off-axis longitudinal pumping in spherical optical cavities. The realization of the FSE theory in optical fields provides abundant possibilities for studies of the fractional-order beam-propagation dynamics. Subsequently, the propagation of beams in FSE with different external potentials and nonlinear terms was investigated [13-17]. In this vein, various soliton states based on FSE in Kerr nonlinear media and lattice potentials were reported recently [18-23]. In particular, it has been found that, with the decrease of LI, solitons become more localized, and their existence region essentially changes [20].

Complex Ginzburg-Landau equations (CGLEs) are universal models for the light propagation in nonlinear dissipative media. They also find a great variety of realizations in other areas, such as superconductivity and superfluidity, fluid dynamics, reaction-diffusion pattern formation, nonlinear optics, Bose–Einstein condensates, quantum-field theories, biology, etc. [23-25]. In particular, CGLEs serve as realistic dynamical models of laser cavities, accounting for the formation of stable fundamental and vortex solitons, as well as multi-soliton clusters [26-37]. The fractional generalization of CGLE was first presented by Weitzner and Zaslavsky [38], and later derived as variational Euler-Lagrange equations in fractal media [39,40]. This model is formulated below in Section 2. Some results for localized wave solutions have been produced by the application of numerical difference schemes to fractional CGLEs [41-47]. However, soliton dynamics has not yet been investigated in various forms of these models.

In this paper, we first develop, in Section 3, an analytical variational approach for solitons in the framework of the fractional nonlinear Schrödinger equation (NLSE), along with an analytical approximation for their exponential tails, in order to explicitly evaluate the shape of such solitons [48]. Numerical results for dissipative solitons of fractional CGLE are reported in Section 4, where the generation and evolution of fundamental solitons and interactions of two solitons are addressed. Numerical results reveal areas of stable soliton propagation, as well as merger of interacting solitons, for different values of the respective LI and dissipation parameters. The paper is concluded by Section 5.

## 2. THE MODEL

We adopt the following natural model of the fractal CGLE type with the cubic-quintic nonlinearity. The CGLE is written in terms of the light propagation along axis *z* in a waveguide with transverse coordinate *x*:

$$iu_z - \frac{1}{2}\left(-\frac{\partial^2}{\partial x^2}\right)^{\alpha/2} u + |u|^2 u - \nu|u|^4 u = iR[u], \quad (1)$$

where the coefficient in front of the cubic self-focusing term is scaled to be 1, $\nu \geq 0$ is the quintic self-defocusing coefficient, and $\alpha$ is the LI belonging to interval $1 < \alpha \leq 2$, as adopted in fractional quantum mechanics and Lévy path integrals [1], as well as in the FSE that occurs in optics [12] (the consideration of the case of $\alpha < 1$ is not relevant, as in this case solitons are unstable because of a possibility of the collapse in the same equation, see Section 3 below). The fractional derivative in Eq. (1) is realized as the integral operator produced by the direct and inverse Fourier transforms [1,12]:

$$\left(-\frac{\partial^2}{\partial x^2}\right)^{\alpha/2} u = \frac{1}{2\pi} \iint dp\, d\xi\, |p|^\alpha \exp[ip(x-\xi)] u(\xi). \quad (2)$$

In the limiting case of $\alpha = 2$, the fractional Laplacian reduces to the classical operator, and Eq. (1) reduces to the commonly known CGLE [23-25]. On the other hand, in the opposite limit of $\alpha = 1$, the operator that seems as a "square root of the Laplacian" appears in a phenomenological model of instability of combustion fronts, see Ref. [49] and references therein.

Loss and gain terms are collected on the right-hand side of Eq. (1),

$$R[u] = -\delta u + \varepsilon|u|^2 u - \mu|u|^4 u + \widehat{D}u, \quad (3)$$

where $\delta > 0$ is the linear-loss coefficient, $\mu > 0$ is the quintic-loss parameter, and

$\varepsilon > 0$ accounts for the cubic gain, which drives dynamics in the model. The last term in Eq. (3) represents an effective diffusion:

$$\widehat{D}u = -\beta(-\frac{\partial^2}{\partial x^2})^{\alpha/2}u, \qquad (4)$$

with the same LI and a positive coefficient, $\beta$, which determines the friction force inhibiting transverse drift of excitations in the medium.

## 3. ANALYTICAL RESULTS: THE VARIATIONAL APPROXIMATION FOR SOLITONS, AND THEIR EXPONENTIAL TAILS, IN THE FRACTAL NONLINEAR SCHRÖDINGER EQUATION

To produce an explicit approximation for solitons in fractal models, we start by the consideration of the basic one in the form of the fractal NLSE with the cubic nonlinearity, i.e., Eq. (1) with $\nu = 0$ and $R[u] = 0$. Stationary soliton solutions with propagation constant $k$ are looked for as $u = \exp(ikz)U(x)$, with real function $U(x)$ satisfying the integro-differential equation,

$$kU(x) + \frac{1}{4\pi}\iint dpd\xi |p|^{\alpha}\exp[ip(x-\xi)]U(\xi) - U^3(x) = 0 \qquad (5a)$$

The Lagrangian from which this equation can be derived is

$$L = \frac{1}{2}k\int_{-\infty}^{+\infty}U^2(x)dx - \frac{1}{4}\int_{-\infty}^{+\infty}U^4(x)dx$$

$$+ \frac{1}{8\pi}\iiint dpdxdx'|p|^{\alpha}\exp[ip(x-x')]U(x')U(x) \qquad (5b)$$

A natural variational ansatz is adopted in the form of a Gaussian [50],

$$U(x) = A\exp(-ax^2/2), \qquad (6)$$

with amplitude $A$ and inverse squared width $a$. The norm of the ansatz is

$$N \equiv \int_{-\infty}^{+\infty}U^2(x)dx = A^2\sqrt{\pi/a}. \qquad (7)$$

The substitution of the ansatz in Lagrangian (5b) yields an expression in which the amplitude is replaced by the norm as per Eq. (7):

$$L = \frac{k}{2}N - \sqrt{\frac{a}{2\pi}}\frac{N^2}{4} + \frac{1}{4\sqrt{\pi}}\Gamma(\frac{\alpha+1}{2})a^{\alpha/2}N, \tag{8}$$

where $\Gamma$ is the gamma-function.

The Lagrangian also determines the model's energy, $E \equiv L - kN/2$. It follows from here and from Eq. (8) that catastrophic self-compression of the soliton at fixed $N$, i.e., $a \to \infty$, leads to $E \to -\infty$, which implies the onset of the *collapse* [51,52] at $\alpha \leq 1$, hence all solitons are unstable in this case. At $\alpha > 1$, the compression leads to $E \to +\infty$, which provides stability of solitons.

Finally, the Euler-Lagrange equations, $\frac{\partial L}{\partial a} = \frac{\partial L}{\partial N} = 0$, yield expressions for $a$ and $k$ in terms of the norm:

$$a = \left(\frac{N}{\sqrt{2}\alpha\Gamma[(\alpha+1)/2]}\right)^{2/(\alpha-1)},$$

$$k = \frac{[1-(2\alpha)^{-1}]}{2^{\alpha/(2(\alpha-1))}\sqrt{\pi}}\{\alpha\Gamma[(\alpha+1)/2]\}^{-1/(\alpha-1)}N^{\alpha/(\alpha-1)}. \tag{9}$$

In particular, in the case of $\alpha = 2$, which corresponds to the usual NLSE with the second-order derivative, Eq. (9) yields the well-known results, $a = \frac{1}{2\pi}N^2$ and $k = \frac{3}{8\pi}N^2$ [50]. These results, along with the form of ansatz (6), supply an explicit approximation for the shape of solitons in the model based on the fractal NLSE.

Note that the $k(N)$ dependence, given by Eq. (9), satisfies the well-known Vakhitov-Kolokolov (VK) criterion, *dk/dN*> 0 [51,52,53], which is a necessary condition for stability of the solitons. It is relevant to mention that scaling $a \propto N^{2/(\alpha-1)}$ and $k \propto N^{\alpha/(\alpha-1)}$, implied by Eq. (9) (and, hence, the validity of the VK criterion) is an *exact property* of Eq. (5a), which is not predicated upon the use of the variational approximation.

Further, we note that, in the limit case of $\alpha = 1$, which, as mentioned above, is a boundary of the collapse regime, relation (9) for the propagation constant becomes *degenerate*, yielding $N = \sqrt{2}$, i.e., the norm may take a single value, which does not depend on *k*. While this particular value is an approximate one, produced by the variational method, the fact that *N* may only assume a single value is an exact property of the *critical collapse* [51,52], which is a boundary separating the *supercritical collapse* at $\alpha < 1$ and collapse-free dynamics at $\alpha > 1$.

Comparison of the variational approximation prediction for the solitons' shape with numerical solutions of Eq. (5a) is presented in Fig. 1. The prediction is basically accurate, with small differences in the amplitude and width. For these examples, the amplitude predicted by Eq. (9) differs by less than 5% from the corresponding numerical value, *U*(*x*=0). Another relatively weak discrepancy is the presence of small-amplitude tails featured by numerically generated profiles (see below), which are not captured by the Gaussian ansatz (6).

Further, Fig. 2 displays the comparison of the dependence *N*(*k*) for the solitons, produced by the numerical solution of the fractal NLSE (5a) with the cubic nonlinearity, and its counterpart predicted by the variational approximation as per Eqs. (7) and (9). The largest relative discrepancy, corresponding to *k* = 6 in Fig. 2, is 5.5%.

The above-mentioned tails of the soliton's shape can be found in an approximate analytical form too. Indeed, the consideration of the linearization of Eq. (5a) (with the cubic term dropped) at |*x*|→∞ demonstrates that the exponentially decaying form of the tail,

$$U(x) = U_0 \exp(-q|x|), \qquad (10)$$

with some $q > 0$, is compatible with the linearized equation. Further, redefining the integration variable in the second term in Eq. (5a) as $\xi \equiv x + \delta\xi$, and applying obvious rescaling, $\delta\xi' \equiv q\delta\xi, p' \equiv p/q$, it is easy to see that the integral term takes the form of $-(1/2)C_\alpha q^\alpha U_0$, where $C_\alpha$ is a real constant that cannot be calculated analytically for arbitrary $\alpha$ (a particular value is $C_{\alpha=2} = 1$). Finally, the substitution of this form of the integral term in the linearized equation (5a) makes it possible to find the dependence of the exponential-decay factor, *q* in Eq. (10), on the propagation constant, *k*: $q = ((2/C_\alpha)k)^{1/\alpha}$.

## 4. NUMERICAL RESULTS

Generic results produced by systematic numerical simulations of Eq. (1) may be adequately represented for parameters fixed as $\nu = 0.115$ and $\mu = 1$. The input is a Gaussian beam(cf. ansatz (6)),

$$u = u_0 \exp\left(-\frac{x^2}{2w^2}\right), \qquad (11)$$

where $u_0$ and $w$ represent the amplitude and width of the input, respectively. The split-step fast-Fourier-transform method was adopted to simulate the evolution of the input in the framework of Eq. (1), with fixed $u_0 = 1.2$ and $w=1.0$.

To report results of numerical simulations, it is essential to monitor the dependence of the outcomes on LI, as it determines the effective fractional dimension of the model. First, the findings are systematically summarized in Fig. 3(a) in the plane of $\alpha$ and linear-loss coefficient $\delta$, for fixed values of the diffusion and cubic-gain coefficients, $\beta=0.1$ and $\varepsilon=1.7$. In region B of Fig. 3(a), the input expands into a spreading pattern, the loss coefficient $\delta$ being too small to support stable propagation, as seen in Fig. 3(b). When $\delta$ takes values in area C between solid and dashed curves in Fig. 3(a), the propagation seems somewhat unstable at an initial stage, developing into a stable soliton-like state [Fig. 3(c)]. As $\delta$ increases further, bringing one into area D in Fig. 3(a), stable solitons readily self-trap [Fig. 3(d)]. Further, in region E of Fig. 3(a), the input decays under the action of the strong loss, as shown in Fig. 3(e). Similar propagation scenarios are revealed by simulations of Eq. (1) for larger values of the diffusion coefficient, such as $\beta=0.2$, the only difference from its counterpart in Fig. 3(a) being a small leftward shift of all boundaries at relatively small values of LI $\alpha$.

Next, we summarize the results in another relevant parameter plane, whose coordinates are the LI $\alpha$ and diffusion coefficient $\beta$. It is found that the parameter region in which stable solitons self-trap narrows in the direction of $\beta$ with the increase of $\alpha$, see region D in Fig. 4. At $\beta>3.98$ for a fixed value of the linear-loss parameter,

$\delta$=0.1, the input rapidly decays, regardless of the value of $\alpha$.

Further, we identify domains of different outcomes of the evolution of the input in the parameter plane of LI and cubic-gain coefficient, $(\alpha, \varepsilon)$. There are two different generic outcomes produced by the simulations in this form, *viz*., decay and self-trapping of stable solitons, shown in Fig. 5. With the growth of $\varepsilon$, the minimum value of $\alpha$ necessary for the stability is decreasing (in other words, a minimum value of the cubic-gain coefficient, above which stable solitons self-trap, increases with the decrease of LI). If other parameters are kept constant, a larger cubic-gain coefficient, $\varepsilon$, is naturally needed to compensate stronger diffusing and larger linear loss.

Finally, we discuss the effect of the fractional dispersion and diffusion operators on interaction of two in-phase solitons, each initially taken in the form of pulse (11). Attraction between them leads to merger, as shown in Fig. 6. In particular, the distance traveled by the pulses before the merger depends on the value of LI $\alpha$, see Fig. 6(a). With the increase of $\alpha$, the merger distance $z$ at first gradually decreases, attains a minimum value, and then gradually increases. A smaller linear-loss coefficient, $\delta$, and larger cubic-gain coefficient, $\varepsilon$, lead to a larger merger distance (see solid, dashed, dotted, and dashed-dotted curves in Fig. 6(a)). Thus, the propagation distance necessary for the merger of in-phase solitons can be controlled by adjusting the LI value.

## 5. CONCLUSIONS

The work addresses two issues for solitons in models based on fractal equations. First, we have derived the variational approximation for solitons of the fractal nonlinear Schrödinger equation, and the asymptotic expression for exponentially decaying solitons' tails. The results demonstrate that the Gaussian ansatz provides a sufficiently accurate approximation for the shape of the solitons. Then, we have studied the evolution of inputs in the form of Gaussians in the fractional complex Ginzburg-Landau equation, which includes fractional operators in both the dispersion and diffusion terms. Effects of the value of the respective Lévy index (LI) on stability regions of dissipative solitons generated by the model have been investigated in detail. It is demonstrated that a pair of in-phase Gaussians merge into a single soliton, with the merger distance also determined by LI.

**Acknowledgments**

This work was supported by the National Natural Science Foundations of China (Grant Nos. 61675001, 11774068, and 11604050), the Guangdong Province Nature Foundation of China (Grant No. 2017A030311025), and the Guangdong Province Education Department Foundation of China (Grant No. 2014KZDXM059).

# Figures

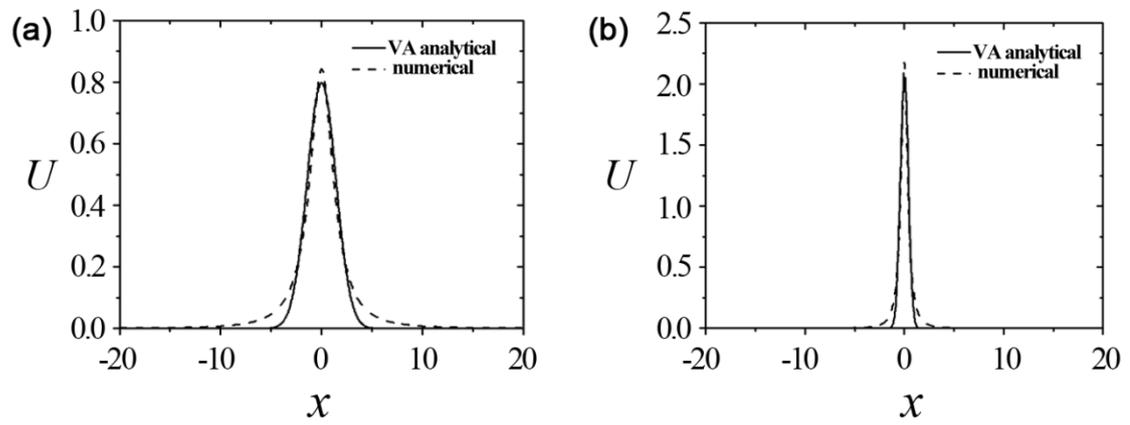

Fig. 1. Comparison of the shape of the solitons predicted by the variational approximation ("VA analytical") with numerical solutions of Eq.(5a) at $\alpha=1.5$ with (a) $k=0.3$ and (b) $k=2.0$.

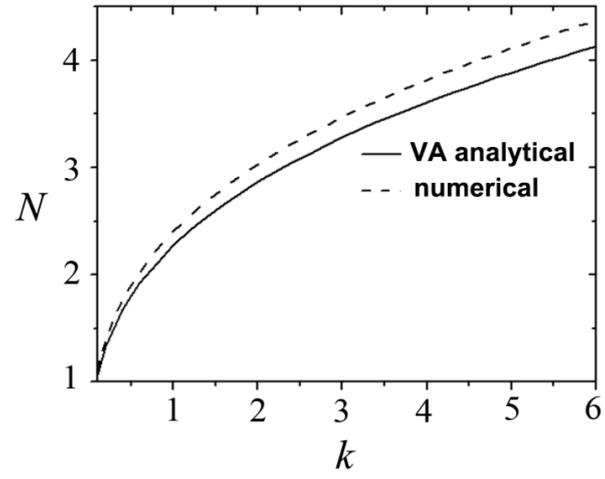

Fig. 2. Dependence *N(k)* for the solitons of Eq. (5a), obtained from the numerical solution and variational approximation at $\alpha=1.5$.

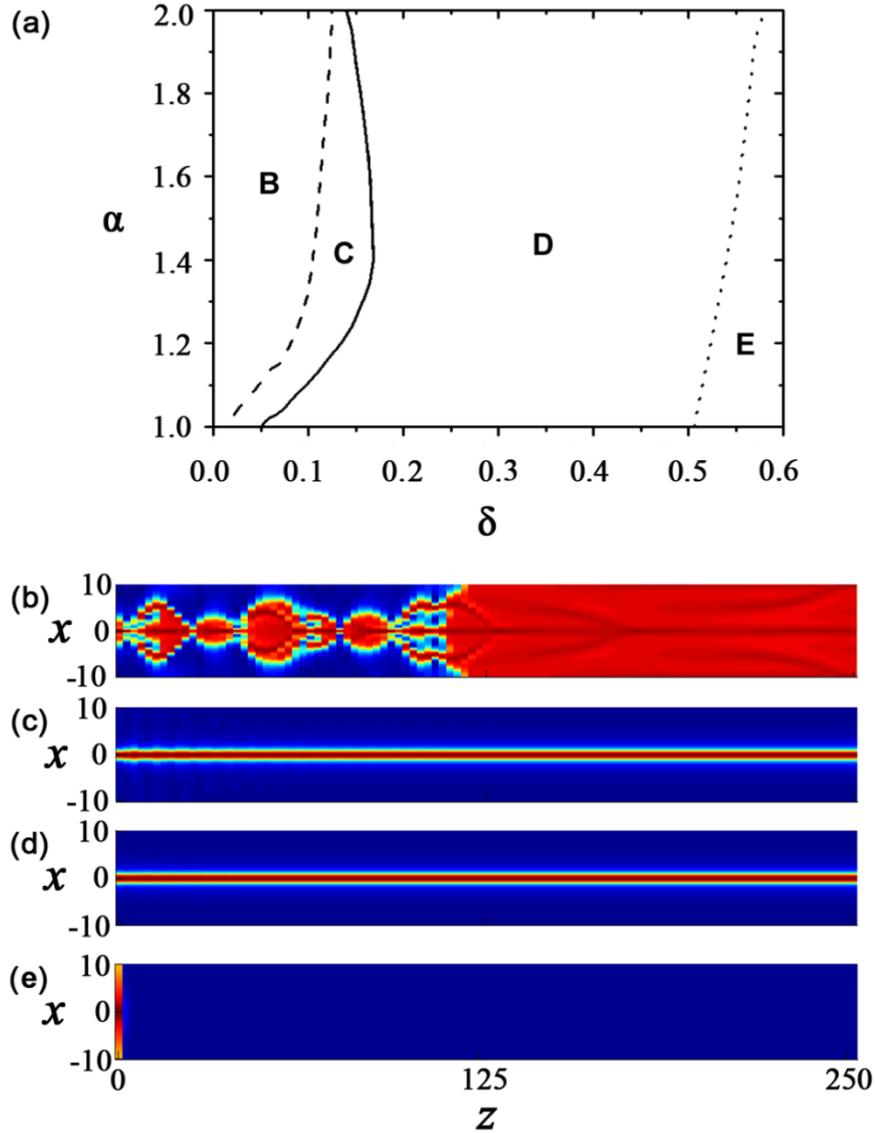

Fig. 3. (Color online) (a) Domains of different propagation scenarios of the input Gaussian beam (11) in the plane of ($\delta,\alpha$), for fixed $\beta=0.1$ and $\varepsilon=1.7$. Region B: expansion of the input in the underdamped setting. Region C: initial unstable evolution followed by self-trapping of a stable soliton. Region D: fast formation of stable solitons. (b) An example of the unstable propagation for $\delta=0.1$ and $\alpha=1.4$, corresponding to region B in panel (a). (c) The original unstable evolution followed by the formation of a stable soliton, for $\delta=0.1$ and $\alpha=1.2$ (region C in panel (a)). (d) Fast formation of a stable soliton for $\delta=0.1$ and $\alpha=1.05$ (corresponding to region D in (a)). (e) Decay of input beam for for $\delta=0.55$ and $\alpha=1.05$, which occurs in region E in panel (a).

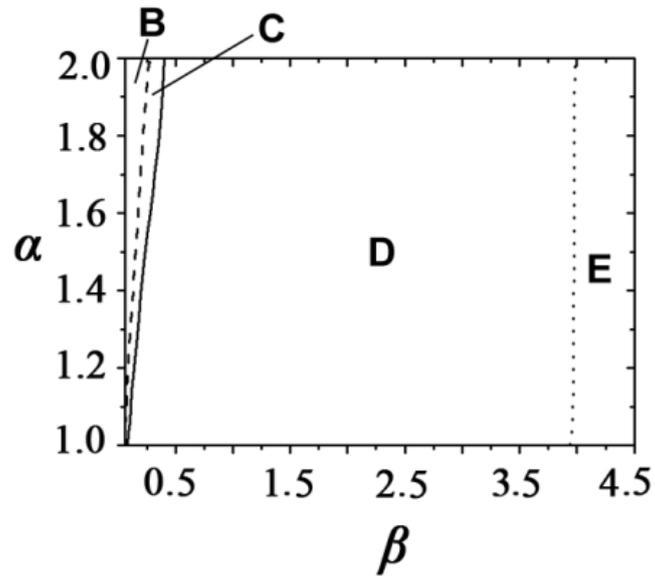

Fig. 4. The same regions B, C, D, and E as shown in Fig.1(a), but in the ($\beta,\alpha$) plane at fixed values of the linear-loss coefficient $\delta=0.1$. Other parameters are the same as in Fig. 3.

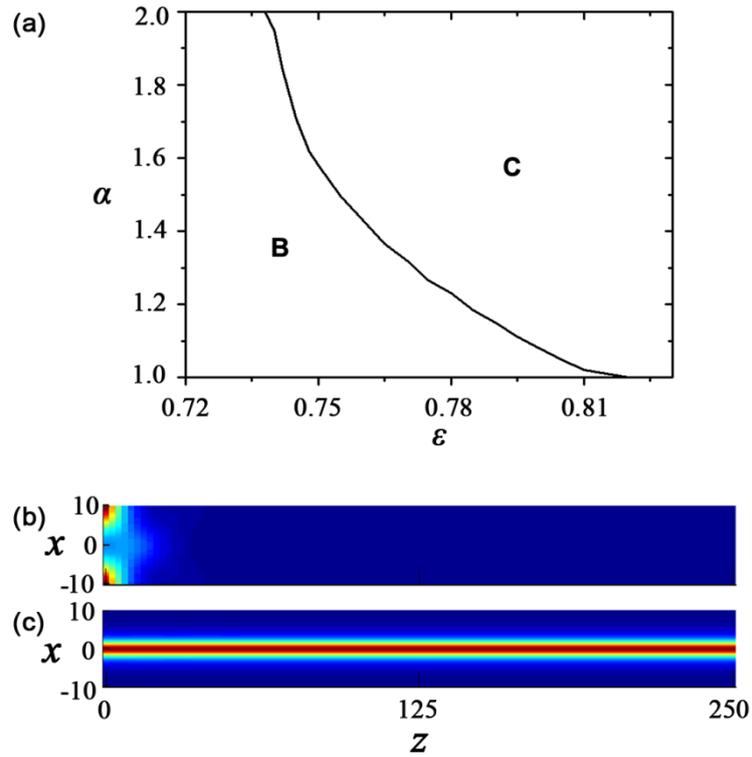

Fig. 5. (Color online) Domains of different propagation scenarios of the Gaussian input (11) in the plane of $(\varepsilon,\alpha)$, for $\delta=0.1$ and $\beta=0.1$ in (a). Region B: decay of the input in the overdamped setting; region C: self-trapping of stable solitons. (b) An example of the decay of the input for $\varepsilon=0.78$ and $\alpha=1.1$, corresponding to region B in panel (a). (c) Self-trapping of a stable soliton for $\varepsilon=0.78$ and $\alpha=1.6$, corresponding to region D in (a).

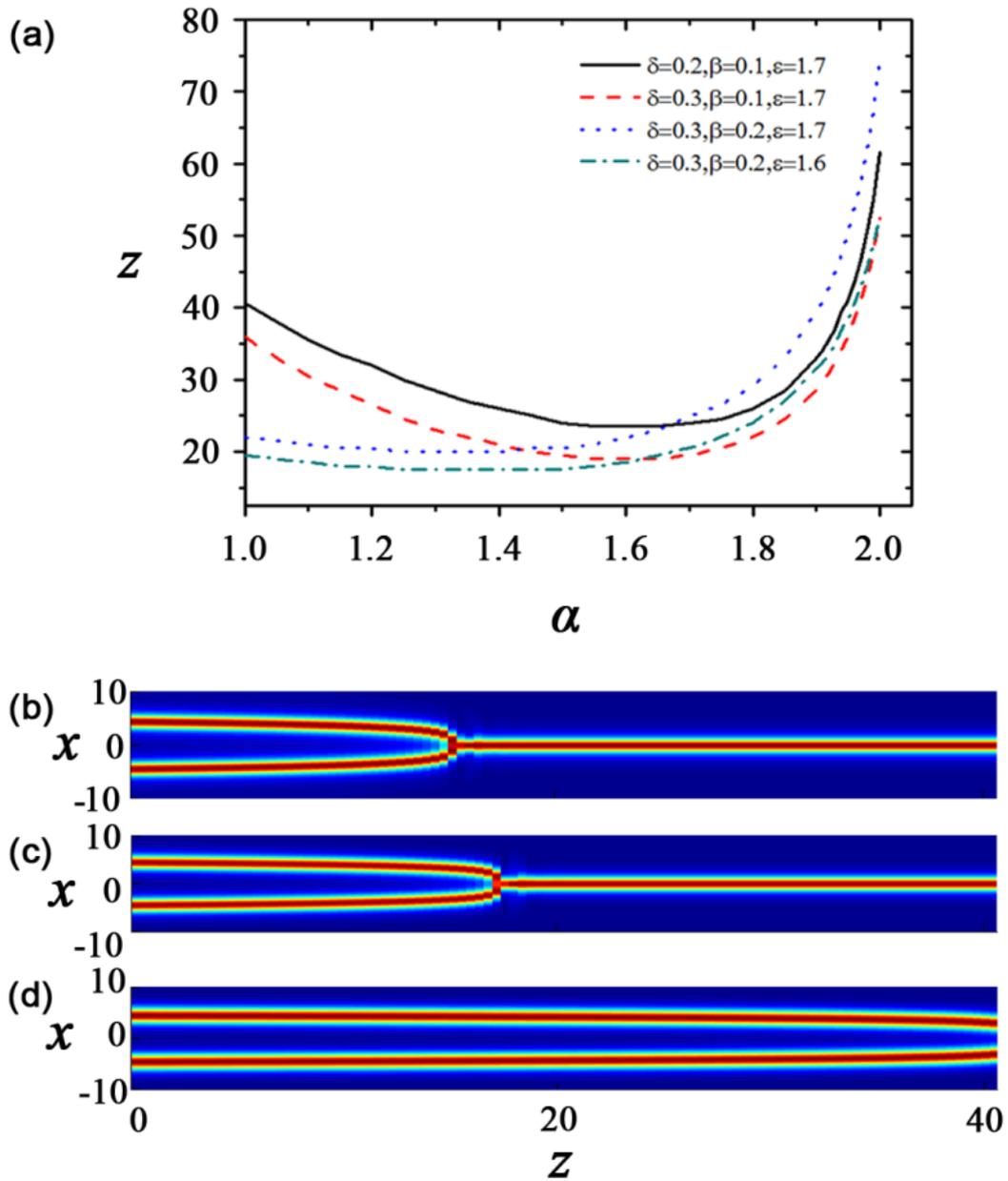

Fig. 6. (Color online) (a) Propagation distance $z$ necessary for the merger of two in-phase Gaussians as a function of LI, $\alpha$. Examples of the merger for fixed $\delta=0.3$, $\beta=0.1$, $\varepsilon = 1.7$ and different values of LI: $\alpha = 1.5$ (b), $\alpha = 1.8$ (c), and $\alpha = 2.0$ (d).